\documentclass[%
 aip,
 amsmath,amssymb,
 reprint,%
]{revtex4-1}

\usepackage{graphicx}
\usepackage{dcolumn}
\usepackage{bm}

\usepackage[utf8]{inputenc}
\usepackage[T1]{fontenc}
\usepackage{mathptmx}
\usepackage{etoolbox}

\usepackage{xcolor} 

\makeatletter
\def\@email#1#2{%
 \endgroup
 \patchcmd{\titleblock@produce}
  {\frontmatter@RRAPformat}
  {\frontmatter@RRAPformat{\produce@RRAP{*#1\href{mailto:#2}{#2}}}\frontmatter@RRAPformat}
  {}{}
}%
\makeatother
\begin{document}

\preprint{AIP/123-QED}

\title[Flow stability and permeability in a nonrandom porous medium analog]{Flow stability and permeability in a nonrandom porous medium analog}
\author{T.P. Le\~{a}o}
\email{tleao@unb.br.}
\affiliation{Soil Physics Laboratory, Faculty of Agronomy and Veterinary Medicine, University of Brasilia. Brasilia, Distrito Federal, Brazil, 70910-900
}


\date{\today}

\begin{abstract}
The estimation of the permeability of porous media to fluids is of fundamental importance in fields as diverse as oil and gas industry, agriculture, hydrology and medicine. Despite more than 150 years since the publication of Darcy's linear law for flow in porous media, several questions remain regarding the range of validity of this law, the constancy of the permeability coefficient and how to define the transition from Darcy flow to other flow regimes. This study is a numerical investigation of the permeability and flow stability in a nonrandom quasi-tridimensional porous medium analog. The effect of increasing pressure gradient on the velocity field and on the estimation of Darcy and Darcy-Forchheimer coefficients is investigated for three different obstacle's radiuses. The transition from Darcy flow to nonlinear behavior is associated with the formation of jets in the outlet of the porous medium and development of flow instabilities. Different representations of the Reynolds number proved adequate to detect deviation from the linear law. The instantaneous permeability calculated at each pressure gradient was sensitive to flow velocity, in agreement with previous studies stating that permeability cannot be conceptualized as a constant for real flows.     
\end{abstract}

\maketitle


\section{\label{sec:level1} Introduction}

Flow in porous media includes a range of processes and models of critical importance in recharge, discharge and exploration of groundwater \cite{doi:https://doi.org/10.1002/9781119300762.wsts0027}, transport of contaminants in the subsurface \cite{bear1988dynamics, *SHU2023606}, oil, gas and CO$_2$ dynamics \cite{WANG2022137957}, geothecnical engineering \cite{https://doi.org/10.1002/nag.2528}, and industrial and research applications \cite{doi:10.1080/10407782.2011.572761}, including medical \cite{KAHSHAN2020223}. Henry Darcy was the first to propose a linear law for the relationship between flow velocity and pressure gradient in porous media, a quadratic term was later added by Forchheimer to account for nonlinear effects as flow velocity is increased \cite{zhang_mehrabian_2021, *ehlers22}. Because of the phenomenological nature of Darcy's law, ever since the development of the framework by Navier and Stokes \cite{batchelor67}, there has been attempts to unify the two equations for porous media by deriving Darcy's law from the Navier-Stokes equations \cite{bear1988dynamics, *barenblatt90}. However, the stochastic nature of most natural porous media precludes a universal closed-form solution.

The validity of Darcy's linear law relies on Stokes flow through the pore space \cite{bear1988dynamics}. Increasing velocity results in increasing contribution of inertial terms resulting in a deviation from the linear law. Besides the quadratic correction of Forchheimer, higher order correction terms have been proposed \cite{mei_auriault_1991, *zolotukhingayubov22}. As the interstitial velocity increases, and depending on the physical characteristic of the fluid, and on some intrinsic characteristic length of the porous medium, flow instability and turbulence may develop \cite{doi:10.1146/annurev-fluid-010719-060317}. Turbulence in porous media is a complex topic because of the coupled stochastic nature of the porous media itself and that of turbulence \cite{delemos2012turbulence, *doi:10.1146/annurev-fluid-010719-060317}.   

The boundary conditions for simulated flow in porous media are usually constrained by theoretical, mathematical and computational limitations, while flow under natural conditions might be driven by a range of heterogeneous conditions. It is not unusual to define boundary conditions exactly at the boundary of the porous domain, especially when modeling groundwater flow \cite{https://doi.org/10.1111/gwat.12893}. However, there are systems in which the boundary conditions might be defined away from the interface between porous domain and free fluid, examples being industrial applications such as filters installed in fluid conducting tubes and at the contacts between groundwater and surface water, as in rivers, lakes or the ocean, or underground caves \cite{doi:https://doi.org/10.1002/9781119300762.wsts0027}. For Stokes flow at low Reynolds number this is known as a Stokes-Darcy system \cite{lyu_wang_2021}. Although the theoretical basis for the boundary conditions between porous media and free fluid when flow is tangential to the interface is described and validated using the Beavers-Joseph interface law \cite{carraroetal2013, *lyu_wang_2021}, many open questions remain for infiltration of the fluid into the porous material \cite{LEVY1975923, *CARRARO2015195, *MARUSICPALOKA2022103638}.  However, even if the boundary conditions can be defined for Stokes-Darcy flow at the inlet and outlet, as the flow velocity is increased, there is a possibility of emergent jets might disturb the flow and boundary conditions at the outlet \cite{PhysRevE.51.5725}, which might become more concerning as the turbulent regime is approached. When the outlet of a porous domain is in contact with a free fluid, the presence of emergent jets can influence the flow and transport characteristics, including stability. Emergent jets can enhance mixing of the fluid and of dissolved contaminants as well as influence heat transport \cite{10.1063/5.0002125, *PAYSANT2021108862}. The presence of emergent jets, or even of nonlinear effects, can also affect flow of fluids into wells, influencing the predictions of output of water, oil, gas and contaminants (see for example \cite{Zeng2006} and references therein).  

This work presents a numerical investigation of flow in a nonrandom porous medium analog composed of cylinders in a quasi-three-dimensional numerical domain. The effects of emergent jets and boundary conditions on flow are investigated based on Darcy and Darcy-Forchheimer calculations using data from direct numerical solutions of the Navier-Stokes equations.

\section{\label{sec:level1} Mathematical models}
Flow in porous media has been modeled using a class of experimentally derived equations first developed by Darcy, written in modern notation as \cite{batchelor67}

\begin{equation}
\langle \mathbf{u} \rangle = - \frac{\boldsymbol{\tilde{\kappa}}}{\mu} \nabla{\langle p \rangle}
\label{eq1}
\end{equation}
where $ \langle \mathbf{u} \rangle$ is the average fluid velocity at the boundaries, called filter velocity in chemical engineering, seepage velocity or flux in groundwater hydrology \cite{scheidegger74, *barenblatt90}, $\mu$ is the dynamic viscosity, $\langle p \rangle $ is the average pressure at the boundaries and $ \boldsymbol{\tilde{\kappa}} $ is a second order tensor representing the fluid permeability of the porous medium, appropriately called the permeability tensor \cite{bear1988dynamics}. 

For an isotropic and homogeneous medium, the permeability tensor reduces to a constant called the permeability $\kappa$, thus   

\begin{equation}
 \langle \mathbf{u} \rangle  = - \frac{\kappa}{\mu} \nabla{\langle p \rangle}
\label{eq2}
\end{equation}
Darcy's law in this form is valid for steady-state flows where the inertial component is much less than the viscous component, as traditionally quantified by the Reynolds number for flow in porous media. There are different interpretations on how the Reynolds number is calculated for porous media \cite{doi:10.1146/annurev-fluid-010719-060317}
\begin{equation}
Re_p =  \frac{\rho \langle \mathbf{u} \rangle d_p}{\mu} 
\label{eq3}
\end{equation}

\begin{equation}
Re_H =  \frac{\rho \langle \mathbf{u} \rangle d_p}{\mu}  \frac{\phi}{1-\phi} 
\label{eq4}
\end{equation}

\begin{equation}
Re_\kappa =  \frac{\rho \langle \mathbf{u} \rangle \sqrt \kappa}{\mu}   
\label{eq5}
\end{equation}
where $d_p$ is some characteristic dimension of the porous medium, usually particle diameter, assumed as equivalent to pore diameter in sphere packing, and $\phi$ is the porosity. The representations of the Reynolds number for porous domains are based on the particle diameter ($Re_p$), hydraulic diameter ($Re_H$) and permeability ($Re_{\kappa}$) \cite{doi:10.1146/annurev-fluid-010719-060317}.  

Where inertial forces become significant, the relationship between flow velocity and pressure gradient can no longer be assumed as linear, thus Darcy's linear law ceases to be valid. A corrected form was presented by Forchheimer and is known as the Darcy-Forchheimer equation \cite{forchheimer14, *zhang_mehrabian_2021}

\begin{equation}
	- \nabla{\langle p \rangle} = \frac{\mu}{\kappa} \langle \mathbf{u} \rangle + \beta \rho \langle \mathbf{u} \rangle \lvert\langle \mathbf{u} \rangle \rvert  
\label{eq6}
\end{equation}
where $\beta$ is the Forchheimer coefficient which quantifies inertial contribution to the pressure gradient \cite{zhang_mehrabian_2021}. The Forchheimer coefficient is also referred to as the ``inertial'' and ``non-Darcy'' coefficient and it is generally agreed to be related to microscopic inertial forces \cite{Ma1993} although an explanation based on microscopic viscous forces has been proposed \cite{Hassanizadeh1987}.  

Following \cite{Ruth1992}, Forchheimer equation can be stated as a general case of the Darcy equation by writing 

\begin{equation}
-  \nabla{\langle p \rangle} =  \langle \mathbf{u} \rangle\frac{\mu}{\kappa} 
\label{eq7}
\end{equation}
in which
\begin{equation}
\frac{\mu}{\kappa}  = \frac{\mu}{\kappa_0} + \beta \rho \langle\mathbf{u}\rangle   
\label{eq8}
\end{equation}
The new coefficient $\kappa_0$ is the permeability as velocity goes to zero, in which case Darcy's equation results. This equation also shows that $\kappa$ is \emph{never} a constant. In Darcy's law the magnitude of the inertial or non-Darcy term $\beta \rho \langle \mathbf{u} \rangle \lvert\langle \mathbf{u} \rangle \rvert  $ is simply much smaller than that of the viscous component $ \langle \mathbf{u} \rangle \mu / \kappa_0 $ so that the latter is neglected. Considering the ratio of inertial to viscous components, \cite{Ruth1992} adapted the corrected Reynolds number proposed by \cite{10.1115/1.4010218}, renaming it the Forchheimer number
\begin{equation}
Fo =  \frac{ \rho \langle \mathbf{u} \rangle \beta \kappa_0}{\mu}   
\label{eq9}
\end{equation}
Later, \cite{Zeng2006} proposed the non-Darcy effect $E$ as a measure of the percent error induced by ignoring non-Darcy effects
\begin{equation}
E =  100\frac{Fo}{1 + Fo}   
\label{eq10}
\end{equation}

Notice that the permeability calculated from the Darcy-Forchheimer equation, referred to in this study as $\kappa_0$, is not the same as that calculated using Darcy's law $\kappa$. The permeability from Darcy's law can also be calculated by fitting Equation \ref{eq2} to a dataset including a range of measured pressure gradients and average velocities using linear least-squares for example, or it can be calculated at a single pressure gradient, as it is often done in laboratory methods in soil physics and hydrology \cite{scheidegger74, *bear1988dynamics}
\begin{equation}
\kappa_D =  \frac{\langle u_x \rangle \mu}{\partial_x p}    
\label{eq11}
\end{equation}
To distinguish between the two, the permeability from Darcy's law calculated using a single pressure gradient will be called instantaneous permeability $\kappa_D$. 

Darcy's law being of empirical nature reduces the internal structure of the porous medium to a stochastic representation, the permeability being a parameter which represents geometric properties of the porous medium \cite{barenblatt90}. The same is true for functions deriving from Darcy's principles such as the Darcy-Forchheimer equation and its parameters $\kappa_0$ and $\beta$.

If the internal structure of the porous medium is known at every point of the domain under consideration, the flow within it can be characterized by a solution of the Navier-Stokes equation. For an incompressible fluid with the absence of body forces, the Navier-Stokes equations is

\begin{equation}
\partial_t \mathbf{u} + (\mathbf{u} \cdot \nabla )\mathbf{u}   = - \frac {\nabla p}{\rho} + \frac{\mu}{\rho}\nabla^2 \mathbf{u}   
\label{eq12}
\end{equation}


\begin{equation}
\text{div} ~\mathbf{u} = 0
\label{eq13}
\end{equation}

Solutions of the Navier-Stokes equation in this form should be able to completely describe the nature of the flow within the porous domain, including flow velocity variations with time, inertial components and flow instabilities. 
There have been several attempts to derive Darcy's law from the Navier-Stokes equations, in essence, most of these derivations rely on some form of averaging of the Navier-Stokes equations, either using simpler techniques such as volume averaging \cite{whitaker86, *slattery99} or more sophisticated such as homogenization theory \cite{beliaevkozlov96, *allaire91}. One problem with this approach is that it must that the average and derived parameters are representative of the whole domain in the scales usually observed in nature. Although often assumed as homogeneous, physical properties of natural porous media are usually heterogeneous and anisotropic, due to natural processes which form sedimentary rocks and deposits, and soils. The second issue is that these derivations are usually based on the Stokes equation and local variability of flow velocity in time, as it is observed in unstable flows, is not accounted for.

Another issue when using Darcy's law pertains to boundary conditions, in many studies fixed pressure or flow velocity boundary conditions are prescribed at the borders of the domain which is often assumed as homogeneous regarding permeability. The transition from porous media to free fluid is not well defined and porous media can cause disturbances in pressure and velocity fields which in turn can influence the estimation of physical parameters. This is a problem from a theoretical point of view, as the question of where a porous medium actually starts is an important one to define simulations and boundary conditions in field and laboratory experiments. It is also an issue in practical applications as these velocity field disturbances can cause preferential flux and mixing of contaminants in transitions from fluid within porous media and free fluid such as in chemical plants, and in the transition of groundwater to free water in rivers, lakes, oceans and wells.   

\section{\label{sec:level1} Numerical simulations}

\begin{figure}[ht]
\centering
 \includegraphics[width=1.0\linewidth]{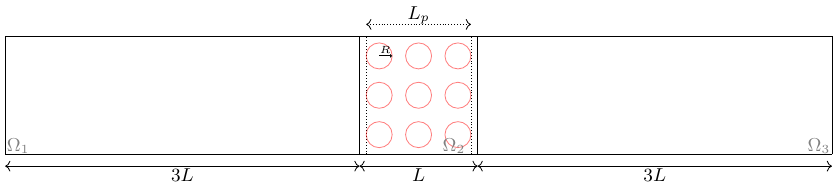}
\caption{Diagram of the system under investigation. }
\label{fig1}
\end{figure}

This study investigates pressure and velocity fields in a simulated cell of a quasi-three-dimensional nonrandom porous medium composed of nine cylinders of equal radius within a channel. One-dimensional permeability is calculated from average velocity and pressure fields at selected lengths in the channel and for varying Reynolds numbers values corresponding to combinations of cylinder radius and flow velocity. The simulation domain $\Omega$ is composed of three subdomains, the free fluid at the inlet $\Omega_1$, the porous medium $\Omega_2$, and the free fluid at the outlet $\Omega_3$ (Figure \ref{fig1}). Because the interfaces between the free fluid and the porous medium are difficult to define, the domain $\Omega_2$ is arbitrarily interpreted as having length $L$ or as having length $L_p$ if the boundary is considered at the limit of the radius of the cylinders. The length of the free fluid domains, $3L$ each, is more than the length reported by \cite{PhysRevE.51.5725}, $2.5L$, aiming to damp the effect of emergent jets on the boundaries. 

Simulations were performed in Open-source Field And Operation Manipulation (OpenFOAM) version 9, using the transient solver for incompressible, laminar flow of Newtonian fluids \emph{icoFoam} with the preconditioned bi-conjugate gradient (PCG) solver with Diagonal Incomplete-Cholesky (DIC) preconditioner for $p$ and \emph{smoothSolver} solver with symmetric Gauss-Seidel \emph{symGaussSeidel} smoother for $u$, with the pressure-implicit split-operator (PISO) algorithm  \cite{greenshields2021}. 

Flow was simulated from left to right in the positive $x$ direction Figure \ref{fig1}. Fluid properties were assumed as dynamic viscosity $\mu = 0.001~ N~ s~ m^{-2}$, density $\rho = 1000 ~kg~ m^{-3}$, thus kinematic viscosity $\nu = \mu/\rho = 0.000001~m^2~s^{-1}$. The dimension $L = 12 \times 10^{-3}~ m $ while the cylinders' radiuses and inlet pressure at the left end of the domain were variables. The radiuses for the cylinders were $0.5\times 10^{-3}~ m $, $1.0\times 10^{-3}~ m $ and $1.5\times 10^{-3}~ m $ while the starting value of kinetic pressure, defined in OpenFoam as the ratio of the pressure by the fluid density, $p_k = p / \rho$, at the inlet (on $\Omega_1$) was $0.00001~ m^2~ s^{-2} $, increased tenfold until the development of flow instabilities for each cylinder radius chosen. The outlet kinematic pressure at the right end of the $\Omega_3$ subdomain was fixed at zero. For the quasi-three-dimensional simulation, the $z$ dimension was fixed at $0.1 \times 10^{-3}~ m $. Upper, lower and cylinder walls were assumed as no-slip boundaries while velocity was set as zero gradient at the inlet and outlet. Grid dimensions were kept constant for each sphere radius at $\delta x = \delta y = 0.1 \times 10^{-3}~ m$ at the boundaries. The end time for each simulation was $5$ with $\delta t = 0.001$. All estimations were performed using data at end time to ensure convergence and steady-state flow. Convergence was monitored using the Courant-Friedrichs-Lewy condition 

\begin{equation}
	Co = \mathbf{u} \frac{\delta t}{\delta x} 
\label{eq14}
\end{equation}

Average pressure and velocity were calculated by slicing the domain in the $y$-$z$ plane and integrating over the area at prescribed points along the $x$ length  
\begin{equation}
	{\langle p \rangle} = \frac {\int p dA}{A} 
\label{eq15}
\end{equation}
    
\begin{equation}
	{\langle u_x \rangle} = \frac {\int \mathbf{u} dA}{A}
\label{eq16}
\end{equation}

Darcy and Darcy-Forchheimer coefficients were calculated from the average velocities and pressures above, considering the gradient $\partial_x p $ over lengths $x$ of interest, $L$ or $L_p$, from Equations \ref{eq11} and \ref{eq6}. The effects of the distance from the porous domain, cylinder radius and pressure gradient on Darcy or Darcy-Forchheimer coefficients estimation were investigated. Reynolds number representations according to equations \ref{eq3}, \ref{eq4}, \ref{eq5} and \ref{eq9} were calculated and compared to Reynolds number for the porous media free portion of the numerical domain  

\begin{equation}
Re_{np} =  \frac{\rho \langle \mathbf{u} \rangle D_H}{\mu}   
\label{eq17}
\end{equation}
in which $D_H$ is the hydraulic diameter

\begin{equation}
D_H =  \frac{2 L_y L_z }{L_y+L_z}   
\label{eq18}
\end{equation}
and $L_y$ and $L_z$ are the $y$ and $z$ dimensions.

\section{\label{sec:level1} Results and discussion}

\subsection{\label{sec:level2}Velocity fields}

\begin{figure}
 \includegraphics[width=1\linewidth]{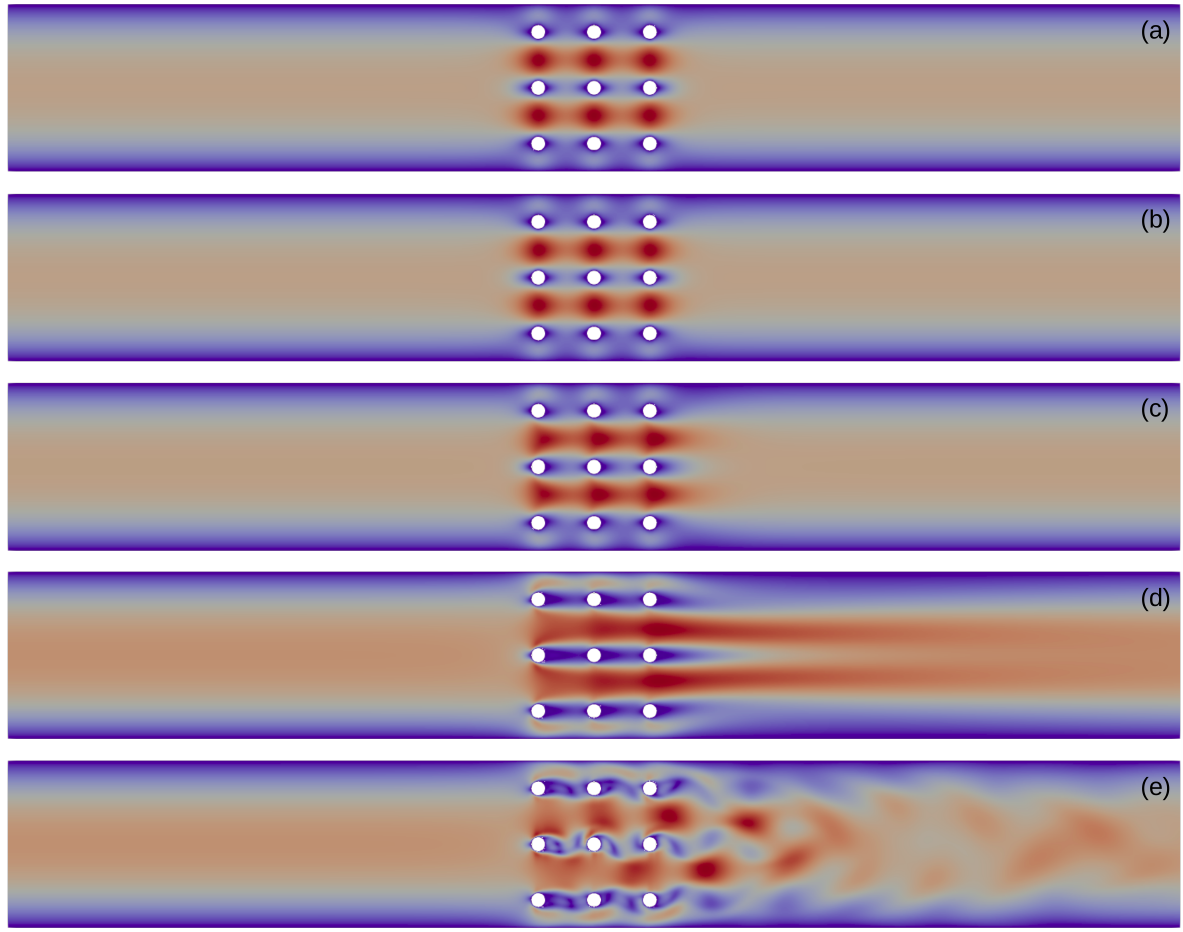}
\caption{Normalized velocity field for $R/L \approx 0.042$. (a) $p_{ki}/p_{kmin} = 1$, (b) $p_{ki}/p_{kmin} = 10$, (c) $p_{ki}/p_{kmin} = 100$, (d) $p_{ki}/p_{kmin} = 1000$ and (e) $p_{ki}/p_{kmin} = 6000$.}
\label{fig2}
\end{figure}

\begin{figure}
 \includegraphics[width=1\linewidth]{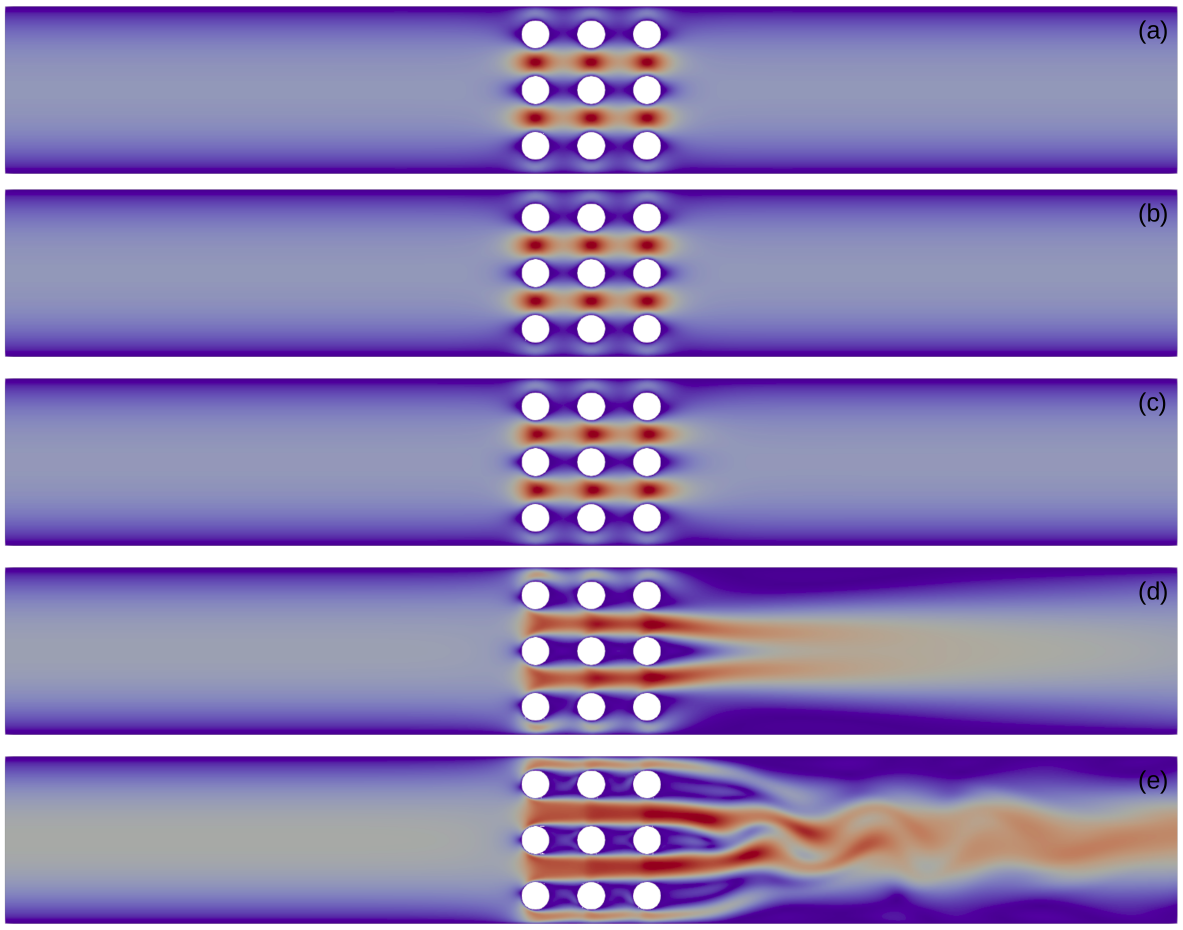}
\caption{Normalized velocity field for $R/L \approx 0.083$. (a) $p_{ki}/p_{kmin} = 1$, (b) $p_{ki}/p_{kmin} = 10$, (c) $p_{ki}/p_{kmin} = 100$, (d) $p_{ki}/p_{kmin} = 1000$ and (e) $p_{ki}/p_{kmin} = 10000$.}
\label{fig3}
\end{figure}

\begin{figure}
\centering
 \includegraphics[width=1\linewidth]{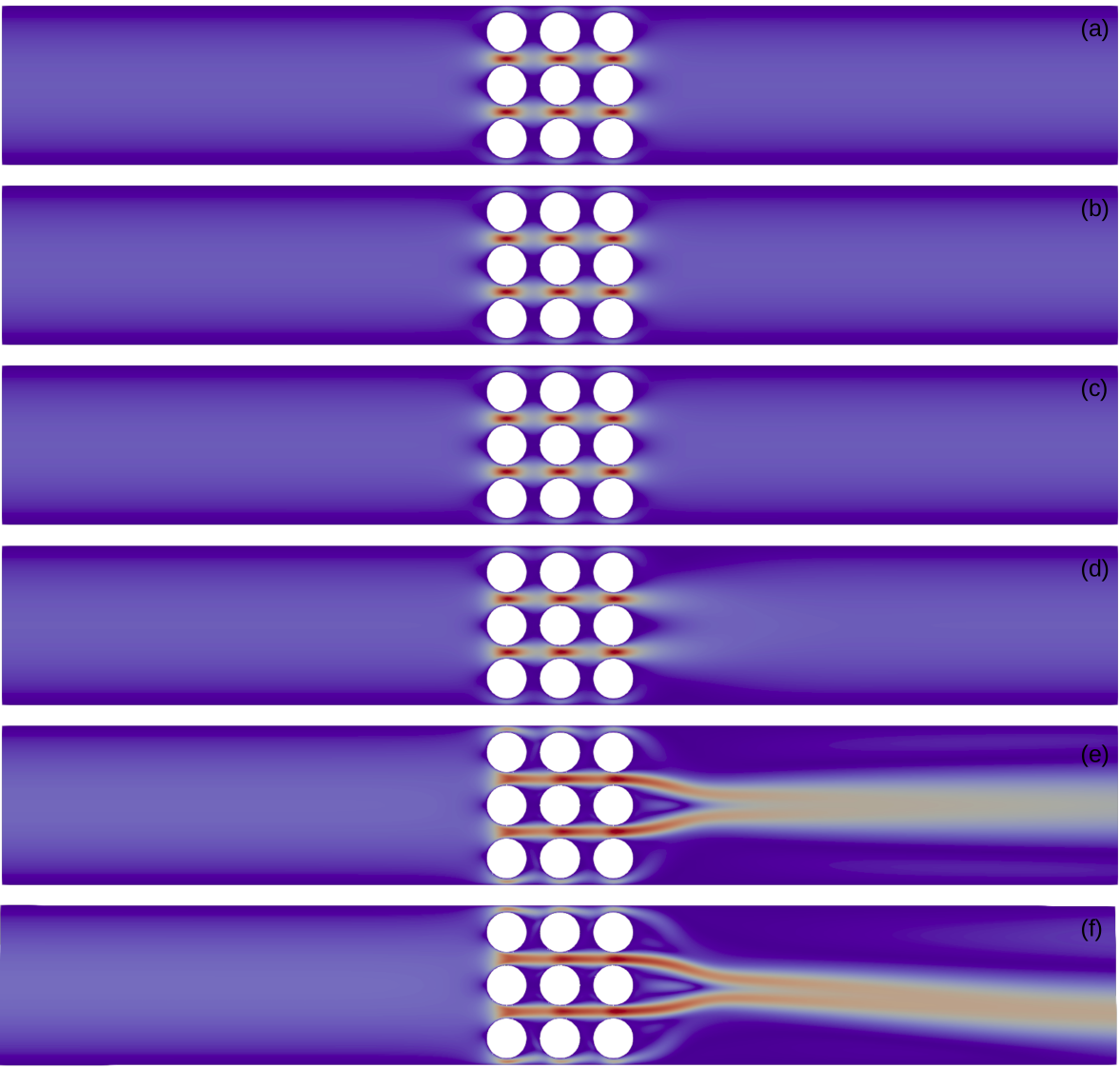}
\caption{Normalized velocity field for $R/L = 0.125$. (a) $p_{ki}/p_{kmin} = 1$, (b) $p_{ki}/p_{kmin} = 10$, (c) $p_{ki}/p_{kmin} = 100$, (d) $p_{ki}/p_{kmin} = 1000$, (e) $p_{ki}/p_{kmin} = 10000$  and  (f) $p_{ki}/p_{kmin} = 20000$.} 
\label{fig4}
\end{figure}

Inspection of the normalized velocity fields with reference to the inlet and outlet boundary between $\Omega_2$ with $\Omega_1$, and $\Omega_2$ with $\Omega_3$ can indicate different regimes of fluid flow. Symmetry between the fields at the inlet and outlet can be associated with low inertia Stokes flow, while development of incipient asymmetric features in the velocity field between inlet and outlet can indicate the appearance of incipient emergent jets and increase in the contribution of inertia. Fully developed emergent jets and flow instabilities characterized by periodic motions of the jets are associated with high inertia and transition to turbulent flows. Symmetry was observed for all cylinder radiuses for low inlet kinetic pressures, early development of asymmetric features being associated with smaller cylinder radius. For $R/L \approx 0.042$  and $R/L \approx 0.083$, symmetry was observed for $p_{ki}/p_{kmin} = 1 $ and $10$ (Figures \ref{fig2}a,b, \ref{fig3}a,b), with $p_{kmin}$ being the lowest values of the inlet kinetic pressure at the left boundary of $\Omega_1$, $1 \times 10^{-6}~m^2~s^{-1}$, used for all simulations and $p_{ki}$ is the actual kinetic pressure at the inlet for each simulation. For $R/L = 0.125$, symmetry was observed for $p_{ki}/p_{kmin} = 1 $, $10$, and $100$ (Figure \ref{fig4}a,b,c). 

Symmetry loss and incipient jet development was observed at $p_{ki}/p_{kmin} = 100$ for $R/L \approx 0.042$  and $R/L \approx 0.083$ (Figure \ref{fig2}c, \ref{fig3}c) and $p_{ki}/p_{kmin} = 1000$ for $R/L = 0.125$ (Figure \ref{fig4}d) and jet formation is evident for $p_{ki}/p_{kmin} = 1000$ for $R/L \approx 0.042$  and $R/L \approx 0.083$ (Figure \ref{fig2}d, \ref{fig3}d) and $p_{ki}/p_{kmin} = 10000$ for $R/L = 0.125$ (Figure \ref{fig4}e). The loss of symmetry and formation of jets is generally associated with boundary layer detachment and concentration of vorticity on one end, in violation of Stokes flow as the inertial components become significant \cite{batchelor67}. 

At higher flow velocities, what is interpreted as jets in the case of periodic porous media can be understood as the wakes after the solid objects, but in this case the wakes after each obstacle are not independent, meaning that the jets interact with each-other, potentially causing mixing and contributing to development of instability. In the case of Figures \ref{fig2}d and \ref{fig3}d, $p_{ki}/p_{kmin} = 1000$, the wake creates asymmetry in the flow field in the $x$ direction, but the flow is still stable within the time frame of the numerical simulation. In both cases the velocity disturbance caused by the wake propagates into the boundary of the subdomain $\Omega_3$, although noticeably damped. For  $R/L \approx 0.042$, the next kinetic pressure for which the $Co$ remained within acceptable limits was $6 \times 10^{-3}~m^2~s^{-1}$, thus $p_{ki}/p_{kmin} = 6000$. In this case the flow became unstable and periodic vortex shedding structures were observed extending into the rightmost boundary, although somewhat damped (Figure \ref{fig2}e). Instability was also observed for $R/L \approx 0.083$, $p_{ki}/p_{kmin} = 10000$, with periodic motion extending into the right boundary, but without fully developed shedding structures (Figure \ref{fig3}e). For $R/L = 0.125$ jet formation is evident for $p_{ki}/p_{kmin} = 10000$, although stable. Instability develops at the maximum acceptable $Co$ investigated, for $p_{ki} = 2 \times 10^{-2}~m^2~s^{-1}$, thus $p_{ki}/p_{kmin} = 20000$. However, instability presented in the form of a slow periodic motion of long frequency, extending into the limit of the domain, captured at the final simulation time in Figure \ref{fig4}f.      

\subsection{\label{sec:level2}Pressure fields averaging}
The ratio of the average kinetic pressure acting in the y-z plane ($\langle p_k \rangle$) at each $x$ over the kinetic pressure at the inlet $(p_{ki})$ was used to investigate the pressure drop along the porous domain and interfaces, and to aid in interpreting the effect of the interface position on the estimate of coefficients (Figure \ref{fig5}). Overall, the increase in kinetic pressure at the $\Omega_1$ inlet increases the pressure gradient at and near the interface. However, the effect becomes more pronounced as the radius of the cylinders decreases. For $R/L \approx 0.042$ and $R/L \approx 0.083$, a pronounced increase pressure gradient near the inlet end of the interface is observed for $p_{ki} /p_{kmin} > 100$ with a corresponding pressure drop at the outlet (Figure \ref{fig5}a,b). For $R/L = 0.125$, the gradient remains approximately constant as flow velocity increases, except for a small pressure drop at the outlet of the porous domain for $p_{ki}/p_{kmin} > 100$ (Figure \ref{fig5}c). 

The increase in the pressure gradient between the porous medium inlet and outlet is consistent with the decrease in symmetry in the velocity fields observed in Figures \ref{fig2}, \ref{fig3} and \ref{fig4}. The low inertia condition creates an approximate symmetry in the velocity field in relation to the midpoint of the porous medium with corresponding anti-symmetry for the pressure fields (Figure \ref{fig5}). As the velocity fields become more asymmetrical, particularly as instabilities begin to develop, the pressure increases near the inlet followed by a sudden drop along the porous domain which continues along the outlet followed by an increase in pressure until reaching the end of the second nonporous domain. In contrast, at low inertia regimes, the pressure decreases almost linearly until reaching the second porous medium interface, followed by a sharp decrease in pressure until reaching the outlet where the pressure continues an approximately linear decrease until reaching the edge of the second nonporous domain. As inlet pressure increases, the anti-symmetry of the pressure fields is gradually lost. 
  
As visualized on the velocity fields, the presence of jets and instabilities is related to increasing pressure gradients between the interfaces. The development of jets and instabilities and higher pressure gradients is related to the decrease in cylinder radiuses. At a given inlet pressure, as the cylinder radius decreases, average flow velocity increases over the entire domain. Thus, in analogy to a classical electrical system, the potential gradient being constant throughout the system, the flow velocity will be restricted by the hydraulic resistance of the porous medium, controlled by the obstacle's radiuses in this case. For larger cylinder radiuses, the volume available for flow decreases, thus increasing the resistance of the system. In a condition where the radius of the cylinders tends to zero, the hydraulic resistance created by the porous medium is null and the flow is the classical Hagen-Poiseuille. The opposite condition would be where the resistance is infinite, the flow velocity is zero and the pressure distribution is a step function across the interface. The resistance analogy is of course not new and has been defined for Darcy flow in porous media \cite{bear1988dynamics} and for turbulent flow in pipes \cite{landau2013fluid}.

\subsection{\label{sec:level2}Permeability estimates}
The decrease in average velocity with increasing $R/L$ ratio can be confirmed from the datasets obtained from the simulations (Tables \ref{table:1}, \ref{table:2} and \ref{table:3}), in agreement with the previous hydraulic resistance discussion. The velocity corresponds to the average in the $x$ direction measured at the inlet of the domain, left side of $\Omega_1$. For each $R/L$ and associated $p_{ki}/ p_{kmin}$, the average velocity $\langle u_{x} \rangle $ and pressure gradient between the inlet and outlet of the porous subdomain $\Omega_2 $ were used to calculate permeability from Darcy's law (Equation \ref{eq11}). For each $R/L$, two conditions for the porous domain were considered, in the first the porous domain has length $L$ in the $x$ direction (Figure \ref{fig1}). In the second condition, the length of the porous domain was considered at the physical limits of the first and last columns of cylinders, thus the length of the porous domain in the $x$ direction was defined as $L_p$ (Figure \ref{fig1}). Considering length $L$ resulted in porosities ($\phi$) of approximately $0.9505$, $0.8037$ and $0.5582$ for $R/L$ of $0.042$, $0.083$ and $0.125$, respectively, while $L_p$ resulted in $0.9346$, $0.7644$ and $0.5181$ for $R/L$ of $0.042$, $0.083$ and $0.125$, respectively.

\begin{table*}
\caption{Conditions for simulation for $R/L \approx 0.042$.}
\centering
\begin{ruledtabular}
\begin{tabular}{c c c c c c c c c c } 
 $ p_{ki}  / p_{kmin}$ & $\langle u_x \rangle$ & $ \partial_x  \langle p_x \rangle$ & $\kappa_D$ & $Re_{np}$ & $Re_{p}$ & $Re_{H}$ & $Re_{\kappa}$ & $Fo$ & $E$\\  
  & $(m~s^{-1})$  & $(Pa~m^{-1})$ & $(m^2)$ &  &  &  & & & $(\%)$\\ 
 \hline
 \multicolumn{10}{c}{\emph{Porous media with length L, $\phi = 0.9509$ }}				\\
 $1$ 	& $2.9993\times 10^{-5}$ & $4.4080\times 10^{-2}$ & $6.8042\times 10^{-7}$ & 0.0059 & 0.0300 & 0.5810 & 0.0247 & 0.0011&0.11 \\ 
 $10$ 	& $2.9985\times 10^{-4}$ & $4.4128\times 10^{-1}$ & $6.7949\times 10^{-7}$ & 0.0595 & 0.2998 & 5.8086 & 0.2472 & 0.0109& 1.07 \\ 
 $100$ 	& $2.9404\times 10^{-3}$ & $4.7807\times 10^{0}$ & $6.1506\times 10^{-7}$ & 0.5832 & 2.9404 & 56.9614 & 2.3060 & 0.1066& 9.63\\ 
 $1000$ & $2.4205\times 10^{-2}$ & $7.4586\times 10^{1}$ & $3.2453\times 10^{-7}$ & 4.8010 & 24.2052 & 468.8984 & 13.7890 & 0.8772& 46.73 \\ 
 $6000$ & $8.6194\times 10^{-2}$ & $5.8273\times 10^{2}$ & $1.4791\times 10^{-7}$ & 17.0964 & 86.1942 & 1669.7390 & 33.1500 & 3.1237 & 75.75 \\ 
 \multicolumn{10}{c}{\emph{Porous media with length L$_p$, $\phi = 0.9346$}}	\\
 $1$ 	& $2.9993\times 10^{-5}$ & $5.8024 \times 10^{-2}$ & $5.1691 \times 10^{-7}$ & 0.0059 & 0.0300 & 0.4283 & 0.0216 &0.0013 & 0.13  \\ 
 $10$ 	& $2.9985\times 10^{-4}$ & $5.8100 \times 10^{-1}$ & $5.1609 \times 10^{-7}$ & 0.0595 & 0.2998 & 4.2815 & 0.2154 & 0.0132 & 1.31  \\ 
 $100$ 	& $2.9404\times 10^{-3}$ & $6.3564 \times 10^{0}$ & $4.6259 \times 10^{-7}$ & 0.5832 & 2.9404 & 41.9860 &  1.9999 &  0.1298 & 11.49  \\ 
 $1000$ & $2.4205\times 10^{-2}$ & $1.0265 \times 10^{1}$ & $2.3580 \times 10^{-7}$ & 4.8010 & 24.2052 & 345.6225 &  11.7537 & 1.0688 & 51.66 \\ 
 $6000$ & $8.6194\times 10^{-2}$ & $8.4857 \times 10^{2}$ & $1.0158 \times 10^{-7}$ & 17.0964 & 86.1942 & 1230.7557 &  27.4709 & 3.8059& 79.19 \\ 
\end{tabular}
\end{ruledtabular}
\label{table:1}
\end{table*}

Calculated permeability was affected by the radius of the cylinders $R/L$, by the choice of length of the porous domain, $L$ or $L_p$, and by flow velocity. Theoretically, Darcy's law is based on the principle that permeability is expected to be constant, an assumption which should hold as long as the flow regime is Stokesian. In other words, Darcy's law is valid for negligible inertia, having no inertial term in its original form. Although permeability was not constant using the numerical simulation data, the range of variability at low flow velocity might be enough to assume that permeability is approximately constant, which is reflected on the dimensionless number discussed in the next section. The permeability decrease between $ p_{ki} / p_{kmin}$  $1$ and $10$ is in the order of $10^{-10}$ for $R/L \approx 0.042$, $10^{-11}$ for $R/L \approx 0.083$  and $10^{-12}$ for $R/L = 0.125 $ and  $10^{-8}$, $10^{-9}$ and $10^{-11}$ between $10$ and $100$ for $R/L \approx 0.042$, $\approx 0.083$ and $0.125$, respectively, increasing one order of magnitude afterwards in all cases. There was no difference in the orders of magnitude of the differences by considering $L$ or $L_p$ for each $R/L$.



\begin{table*}
\caption{Conditions for simulation for $R/L \approx 0.083$.}
\centering
\begin{ruledtabular}
\begin{tabular}{c c c c c c c c c c} 
 $ p_{ki} / p_{kmin}$ & $\langle u_x \rangle$ & $ \partial_x  \langle p_x \rangle$ & $\kappa_D$ & $Re_{np}$ & $Re_{p}$ & $Re_{H}$ & $Re_{\kappa}$ & $Fo$ & $E$ \\  
  & $(m~s^{-1})$  & $(Pa~m^{-1})$ & $(m^2)$ &  &  &  & & & $(\%)$\\ 
 \hline
 \multicolumn{10}{c}{\emph{Porous media with length L, $\phi = 0.8037$ }}				\\
 $1$ & $1.6059\times 10^{-5}$ & $6.9773\times 10^{-2}$ & $2.3016\times 10^{-7}$ & 0.0032 & 0.0321 & 0.1315 & 0.0077 & 0.0007 & 0.07\\ 
 $10$ & $1.6056\times 10^{-4}$ & $6.9778\times 10^{-1}$ & $2.3010\times 10^{-7}$ & 0.0318 & 0.3211 & 1.3143 & 0.0770 & 0.0073& 0.72 \\ 
 $100$ & $1.5801\times 10^{-3}$ & $7.1007\times 10^{0}$ & $2.2253\times 10^{-7}$ & 0.3134 & 3.1602 & 12.9346 & 0.7454 & 0.0716 & 6.68 \\ 
 $1000$ & $1.2194\times 10^{-2}$ & $8.9686\times 10^{1}$ & $1.3596\times 10^{-7}$ & 2.4186 & 24.3872 & 99.8157 & 4.4961 & 0.5525 & 35.59 \\ 
 $10000$ & $5.9289\times 10^{-2}$ & $1.0338\times 10^{3}$ & $5.7350\times 10^{-8}$ & 11.7598 & 118.5782 & 485.3355 & 14.1984 & 2.6865 & 72.87\\ 
 \multicolumn{10}{c}{\emph{Porous media with length L$_p$, $\phi = 0.7644$}}	\\
 $1$ & $1.6059\times 10^{-5}$ & $8.3527\times 10^{-2}$ & $1.9226\times 10^{-7}$ & 0.0032 & 0.1042 & 0.1042 & 0.0070 & 0.0007& 0.07 \\ 
 $10$ & $1.6056\times 10^{-4}$ & $8.3536\times 10^{-1}$ & $1.9220\times 10^{-7}$ & 0.0318 & 0.3211 & 1.0417 & 0.0704 & 0.0070& 0.70 \\ 
 $100$ & $1.5801\times 10^{-3}$ & $8.5207\times 10^{0}$ & $1.8544\times 10^{-7}$ & 0.3134 & 3.1602 & 10.2521 & 0.6804 & 0.0693& 6.48 \\ 
 $1000$ & $1.2194\times 10^{-2}$ & $1.0839\times 10^{2}$ & $1.1250\times 10^{-7}$ & 2.4186 & 24.3872 & 79.1152 & 4.0899 & 0.5351& 34.86 \\ 
 $10000$ & $5.9289\times 10^{-2}$ & $1.2342\times 10^{3}$ & $4.8038\times 10^{-8}$ & 11.7598 & 118.5782 & 384.6832 & 12.9947 & 2.6017 & 72.24\\ 
\end{tabular}
\end{ruledtabular}
\label{table:2}
\end{table*}

\begin{figure}
\centering
 \includegraphics[width=0.75\linewidth]{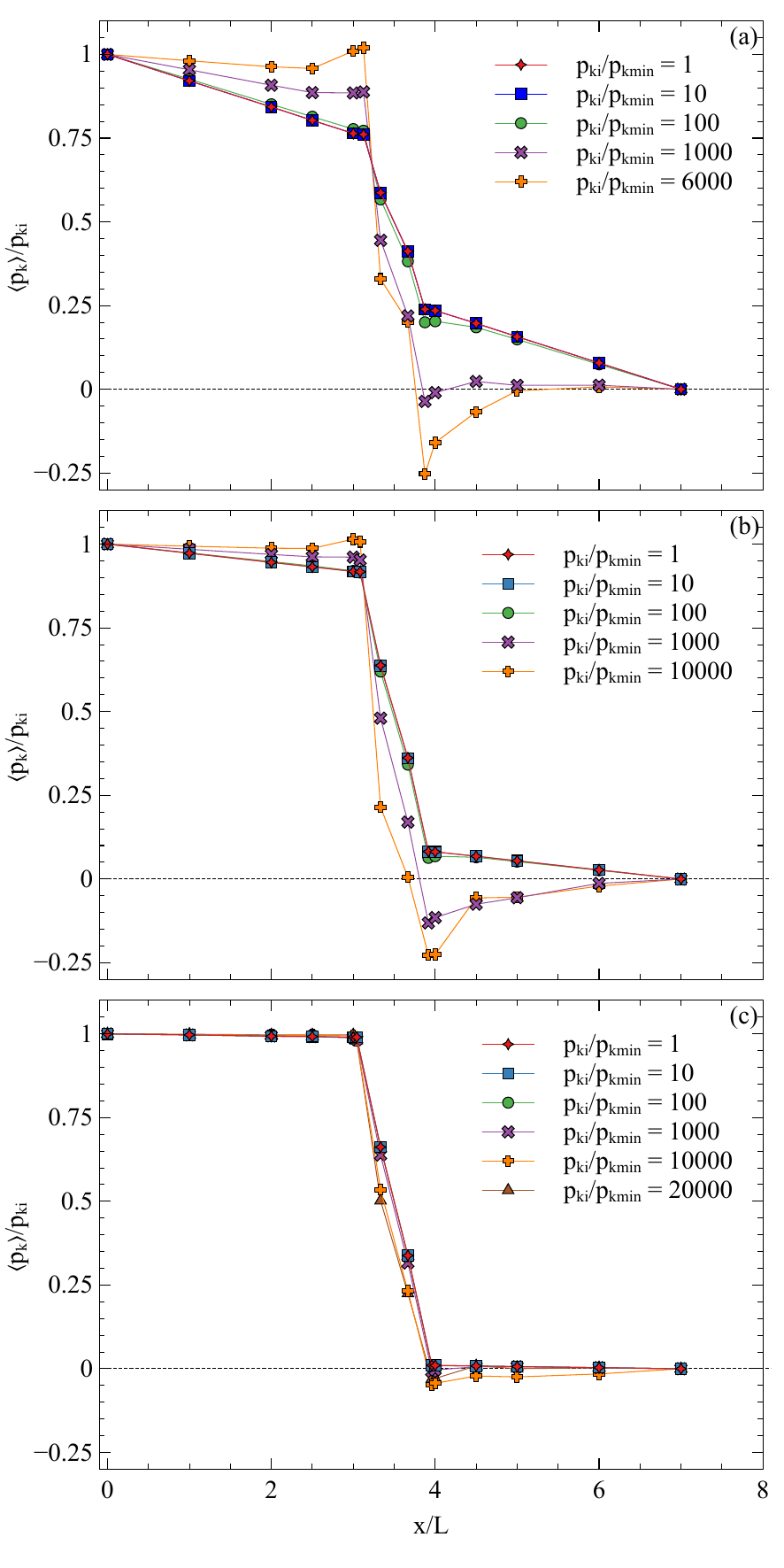}
\caption{Ratio of average kinetic pressure over kinetic pressure at the inlet as a function of relative distance $x/L$. (a) $R/L \approx 0.042$, (b) $R/L \approx 0.083$ and (c) $R/L = 0.125$.}
\label{fig5}
\end{figure}

The choice of length of the porous domain becomes less important for calculation of permeability as flow velocity decreases, or conversely, as obstacle's radius increases. The explanation is that average pressures remain approximatelly constant near the interface at low flow velocities, with little difference in the gradients calculated from $L$ and $L_p$ (Figure \ref{fig5}c). The development of instabilities and jets is associated with a high variability of kinetic pressure near the interface and thus high variability in average flow velocity. Another point is that the range of the instabilities as evaluated by the kinetic pressure averages and normalized velocity fields might extend well into the rightmost limit of the second nonporous domain, thus in extreme cases, a buffer zone corresponding to $3L$ might not suffice to mitigate the effect of jets. However, for Darcian flow, the effect of these disturbances is small, justifying the choice of boundary conditions at the interface or at other arbitrary lengths.  

\begin{table*}
\caption{Conditions for simulation for $R/L = 0.125$.}
\centering
\begin{ruledtabular}
\begin{tabular}{c c c c c c c c c c} 
 $ p_{ki} / p_{kmin}$ & $\langle u_x \rangle$ & $ \partial_x  \langle p_x \rangle$ & $\kappa_D$ & $Re_{np}$ & $Re_{p}$ & $Re_{H}$ & $Re_{\kappa}$  & $Fo$ & $E$ \\  
  & $(m~s^{-1})$  & $(Pa~m^{-1})$ & $(m^2)$ &  &  &  &  &  & $(\%)$  \\ 
 \hline
 \multicolumn{10}{c}{\emph{Porous media with length L, $\phi = 0.5582 $ }}				\\
 $1$ & $3.1291\times 10^{-6}$ & $8.1651\times 10^{-2}$ & $3.8323\times 10^{-8}$ & 0.0006 & 0.0094 & 0.0119 & 0.0006 & 0.0001& 0.01 \\ 
 $10$ & $3.1289\times 10^{-5}$ & $8.1650\times 10^{-1}$ & $3.8321\times 10^{-8}$ & 0.0062 & 0.0939 & 0.1186 & 0.0061 & 0.0013& 0.13\\ 
 $100$ & $3.1268\times 10^{-4}$ & $8.1660\times 10^{0}$ & $3.8291\times 10^{-8}$ & 0.0620 & 0.9380 & 1.1853 & 0.0612 & 0.0134& 1.32 \\ 
 $1000$ & $2.9600\times 10^{-3}$ & $8.2637\times 10^{1}$ & $3.5819\times 10^{-8}$ & 0.5871 & 8.8800 & 11.2202 & 0.5602 & 0.1268 & 11.25 \\ 
 $10000$ & $1.8076\times 10^{-2}$ & $8.6436\times 10^{2}$ & $2.0912\times 10^{-8}$ & 3.5853 & 54.2273 & 68.5181 & 2.6139 & 0.7743& 43.64 \\ 
 $20000$ & $2.8610\times 10^{-2}$ & $1.7095\times 10^{3}$ & $1.6736\times 10^{-8}$ & 5.6748 & 85.8308 & 108.4503 & 3.7013 & 1.2255& 55.07\\ 
 \multicolumn{10}{c}{\emph{Porous media with length L$_p$, $\phi = 0.5181 $}}	\\
 $1$ & $3.1291\times 10^{-6}$ & $8.9033\times 10^{-2}$ & $3.5145\times 10^{-8}$ & 0.0006 & 0.0094 & 0.0101 & 0.0006 & 0.0001 & 0.01 \\ 
 $10$ & $3.1289\times 10^{-5}$ & $8.9033\times 10^{-1}$ & $3.5143\times 10^{-8}$ & 0.0062 & 0.0939 & 0.1009 & 0.0059  & 0.0013 & 0.13 \\ 
 $100$ & $3.1268\times 10^{-4}$ & $8.9044\times 10^{0}$ & $3.5116\times 10^{-8}$ & 0.0620 & 0.9380 & 1.0083 & 0.0586  & 0.0129 & 1.27 \\ 
 $1000$ & $2.9600\times 10^{-3}$ & $9.0053\times 10^{1}$ & $3.2870\times 10^{-8}$ & 0.5871 & 8.8800 & 9.5452 & 0.5366 & 0.1221 &  10.89\\ 
 $10000$ & $1.8076\times 10^{-2}$ & $9.3356\times 10^{2}$ & $1.9362\times 10^{-8}$ & 3.5853 & 54.2273 & 58.2894 & 2.5152  & 0.7459& 42.72\\ 
 $20000$ & $2.8610\times 10^{-2}$ & $1.8381\times 10^{3}$ & $1.5565\times 10^{-8}$ & 5.6748 & 85.8308 & 92.2602 & 3.5694 & 1.1806 & 54.14\\ 
\end{tabular}
\end{ruledtabular}
\label{table:3}
\end{table*}

\subsection{\label{sec:level2}Reynolds number}
Different forms of Reynolds number were calculated and compared to the value for a free channel $Re_{np}$ (Equation \ref{eq12}), namely the equations based on particle diameter $Re_p$ (Equation \ref{eq3}), hydraulic diameter $Re_H$ (Equation \ref{eq4}) and permeability $Re_{\kappa}$ (Equation \ref{eq5}). For $R/L \approx 0.042$, $R_p$ was about 5 times greater than $R_{np}$, while $Re_H$ was about 98 ($L$) and 72 ($L_p$) times greater than $R_{np}$, the ratio between $R_{\kappa}$ and $R_{np}$ was not constant, ranging from around 4 to 2 (for both $L$ and $L_p$)  as inlet pressure increased (Table \ref{table:1}). For $R/L \approx 0.083$, $R_p$ was about 10 times greater than $R_{np}$, while $Re_H$ was about 41 ($L$) and 33 ($L_p$) times greater than $R_{np}$, the ratio between $R_{\kappa}$ and $R_{np}$ ranged from around 2 to 1 (for both $L$ and $L_p$) as inlet pressure increased (Table \ref{table:2}). For $R/L \approx 0.125$, $R_p$ was about 15 times greater than $R_{np}$, while $Re_H$ was about 19 ($L$) and 16 ($L_p$) times greater than $R_{np}$, the ratio between $R_{\kappa}$ and $R_{np}$ ranged from around 1 to 0.6 (for both $L$ and $L_p$)  with increasing inlet pressure (Table \ref{table:3}).


 The constancy between the Reynolds number for the free fluid and $Re_p$ and $Re_H$ is to be expected as they are all calculated from the same average velocity and from static porous medium parameters. $Re_H$ is simply $Re_p$ with a correction for the ratio of volume of pores to that of solids. The variability in the ratio of $Re_{\kappa}$ to the other forms of Reynolds number is because it is corrected to the square root of the permeability. Under low inertia flow conditions, the ratio remains approximately constant, even if the flow velocity is varied. As flow velocity increases and the permeability variation increases, the ratio then starts to vary. Thus, $Re_{\kappa}$ is yet another alternative for measuring the validity of Darcy's law in flow experiments.

As flow velocity decreases, with the increase of cylinder radiuses, the difference between the different methods of calculation of the Reynolds number decreases. At higher flow velocities, in this case $R/L \approx 0.042$, $Re_p$ was closer to $R_{np}$, thus justifying its use for evaluating flow stability. $Re_{H}$ on the other hand was much larger than the other forms at high flow velocities. The use of $Re_{\kappa}$ for measuring flow regime is constrained to the fact that $\kappa_D$ itself if affected by flow velocity and should be used with caution.  However, $Re_{\kappa}$ remained remarkably close to $Re_{np}$ in all cases. 

Based on the previous analysis, the loss of symmetry, development of jets and instabilities and variation in permeability associated with these phenomena appears to be associated with values of $Re_p$ around one thus in agreement with the inertial forces starting to overcome viscous forces as described by the classical Reynolds number. However, $Re_{np}$ and $Re_p$ were not sensitive to the length of the porous domain and, correspondingly, on porosity defined by the length of the porous domain. In this study $Re_\kappa$ was calculated from the instantaneous permeability $\kappa_D$, however, difficulties in defining the appropriate permeability for its calculation might be an obstacle to its use \cite{doi:10.1146/annurev-fluid-010719-060317}.

\subsection{\label{sec:level2}Darcy-Forchheimer coefficients}
\begin{figure}
\centering
 \includegraphics[width=0.75\linewidth]{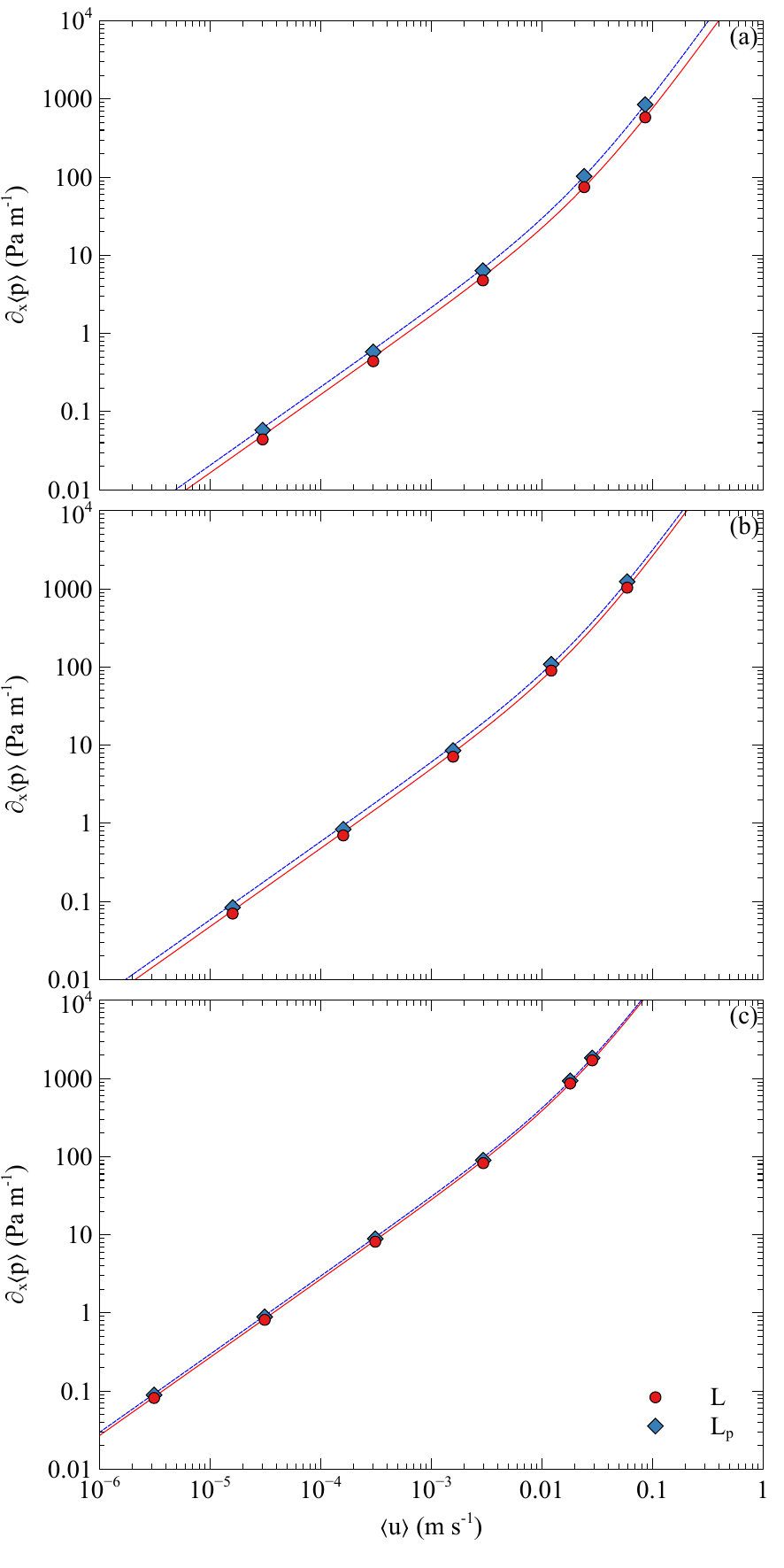}
\caption{Pressure gradient at the porous media versus average inlet velocity. (a) $R/L \approx 0.042$, (b) $R/L \approx 0.083$ and (c) $R/L = 0.125$. Lines are the predictions from Eq. \ref{eq6} with coefficients from Table \ref{table:4}. See text for descriptions of $L$ and $L_p$.}
\label{fig6}
\end{figure}

Darcy-Forchheimer coefficients and associated standard error of the estimates were calculated considering average kinetic pressure gradients for $L$ and $L_p$ by linear least-squares fit of Equation \ref{eq7} (Figure \ref{fig6}). Estimates of $\mu \kappa_0^{-1}$ and $\beta \rho$ as fitting parameters were used to calculate $\kappa_0$ and $\beta$ using $\mu$ and $\rho$ defined for the simulations (Table \ref{table:4}). Instantaneous permeabilities $\kappa_{D}$ calculated directly using Darcy's law at the lowest inlet kinetic pressure $p_{ki}/p_{kmin} = 1$ for each $R/L$ and porous media length (Tables \ref{table:1}, \ref{table:2} and \ref{table:3}) were always higher than $\kappa_0$ estimated using the Darcy-Forchheimer equation. This indicates that direct calculation of permeability at a single pressure gradient might induce systematic errors, as $\kappa_0$ calculated from the Darcy-Forchheimer equation should correspond to the highest permeability values possible when inertial effects, which tend to reduce permeability, are negligible. Another interpretation, however, is that accounting for inertial effects at low velocities and pressure gradients by using the Darcy-Forchheimer equation can induce underestimation of zero velocity permeability and that calculations using Darcy's law are the adequate form at low velocities. This would imply that the true zero velocity permeability is higher than predicted from the Darcy-Forchheimer equation.

\begin{table*}
\caption{Darcy-Forchheimer coefficients. }

\centering
\begin{ruledtabular}
\begin{tabular}{c c c c c} 
 $R/L$  & $ \mu \kappa_0^{-1}  $ & $ \beta\rho  $ & $\kappa_0$ & $\beta$  \\  
  & $(kg~m^{-3}~s^{-1})$  & $(kg~m^{-4})$ & $(m^2)$ & $(m^{-1})$ \\ 
 \hline
\multicolumn{5}{c}{\emph{Porous media with length L}}				\\
 $\approx 0.042$ & $1639.4653\pm 18.5149$ & $59415.4761 \pm 222.5445$ & $6.0995\times 10 ^{-7}$ & $5.9415\times 10 ^{1}$  \\ 
 $\approx 0.083$ & $4729.9038 \pm 54.3304 $ & $ 214323.1437 \pm 935.0298 $ & $2.1142\times 10 ^{-7}$ & $2.1432\times 10 ^{2}$ \\ 
 $0.125 $ & $26864.5368 \pm 563.7848 $ & $1150712.2896\pm 21730.7193$ & $3.7224\times 10 ^{-8}$ & $1.1507\times 10 ^{3}$ \\ 
 \multicolumn{5}{c}{\emph{Porous media with length L$_p$}}	\\
 $\approx 0.042$ & $2048.4956 \pm 15.0038 $ & $90451.7656 \pm 18.3411 $ & $4.8816\times 10 ^{-7}$ & $9.0452\times 10 ^{1}$ \\ 
 $\approx 0.083$ & $5779.7410 \pm 74.5228 $ & $253626.3472 \pm 1282.5419 $ & $1.7302\times 10 ^{-7}$ & $2.5363\times 10 ^{2}$ \\ 
 $0.125$ & $29481.2515 \pm 637.4768 $ & $ 1216592.8735 \pm 24571.1323 $ & $3.3920\times 10 ^{-8}$ & $1.2166\times 10 ^{3}$ \\ 
\end{tabular}
\end{ruledtabular}
\label{table:4}
\end{table*}

Permeability calculated from Darcy's law in all scenarios indicates that it varies with flow velocity. Estimates are higher than those predicted using the coefficients from Table  \ref{table:4} and Equation \ref{eq8} at low velocity but at the higher flow velocities the predictions are very close. Percent error $100 (\kappa_D - \kappa_0)/\kappa_D$ ranged from $10.6\%$ to $4.8\%$ at the lowest velocity to $-0.2\%$ to $-0.4\%$ at the highest velocity considering all simulations. Forchheimer number and non-Darcy effect are presented along with the various representation of Reynolds number in Tables \ref{table:1}, \ref{table:2} and \ref{table:3}. Zeng and Grigg \cite{Zeng2006} state that the non-Darcy error $E$ can be interpreted as the percent error induced by neglecting non-Darcy behavior, stating a cut-off value of $10\%$ for neglecting inertial effects, corresponding to $Fo = 0.11$. Results for $Fo$ and $E$ are agree with the visual interpretation of the symmetry of the velocity fields and the other Reynolds numbers presented regarding transition to non-Darcy flow. However, the discrepancy in the permeability results might indicate the alternative interpretation mentioned previously, that accounting for non-Darcy effects at low velocities might cause underestimation of permeability.   

The value of the Forchheimer coefficient $\beta$ increased with obstacle's radius (Table \ref{table:4}) while there was a decrease in $\kappa_0$. It is difficult to separate these effects from the velocity decrease caused by the increase in obstacle's radius. However, the increase in obstacle's radius increases the surface area available for fluid interactions and energy dissipation. According to \cite{Zeng2006} $\beta \rho \langle u \rangle^2$ can be interpreted as the pressure gradient to overcome liquid-solid interactions while $\mu \langle u \rangle / \kappa_0 $ is the pressure gradient to overcome viscous resistance, thus as surface area increases, $\beta$ should increase correspondingly. However, this might be incompatible with the hypothesis that non-Darcy effects are caused by microscopic inertial forces, on the other hand, the increase in obstacle's radius will also increase the length and tortuosity of flow paths, thus contributing to inertial effects \cite{Ruth1992}. The choice of porous domain length caused reduction of $\kappa_0$ and increase in $\beta$ as length was reduced from $L$ to $L_p$ (Table \ref{table:4}). However, the effect causing difference between $L$ and $L_p$ predictions decrease as obstacle's radius increase with consequent decrease in velocities (Figure \ref{fig6}). This seems to indicate that, as non-Darcy effects become important, the choice of boundary conditions for the porous domain seems to become more critical, in agreement with the symmetry discussion. The analysis of the pressure plots also indicates that the increase in non-Darcy effects increase the pressure gradient along the porous medium, thus contribution to inertial effects (Figure \ref{fig5}). 

\section{\label{sec:level1} Conclusions}
All quantitative indicators show that the development of jets, even if incipient, are associated with the presence of non-Darcy effects, restricting the use of the linear Darcy equation. The permeability in porous media is not constant, not even within the range of validity of Darcy's law. However, within the range of validity of Darcy's law, the order of magnitude of inertial and other potential non-Darcy effects is much smaller than that of the linear effects. The ratio of the inertial, or non-Darcy, to viscous forces for flow in porous media can be expressed using the Forchheimer number and the error induced by ignoring non-Darcy effects by the non-Darcy error. However, most other expressions of the Reynolds number were also adequate to detect transitions in flow regime due to non-Darcy effects, what remains is a matter of critical values. In this numerical investigation, permeability values calculated using Darcy's law indicate that   
values of approximately $Re_{np} \le 0.1$, $Re_p \le 1$, $Re_H \le 10$,  $Re_{\kappa} \le 10$, $Fo \le 0.1$ and $E \le 10\%$ are associated with stokes flow within and outside the porous medium and that non-Darcy effects can be generally neglected. However, more numerical and experimental studies are needed to investigate the discrepancy in permeability values calculated using Darcy and Darcy-Forchheimer equations at low flow velocities.


\section*{Data Availability Statement}

Data available on request from the authors. The data that support the findings of this study are available from the corresponding author upon reasonable request.

\bibliography{aipsamp_nrpm}

\end{document}